# RESULTS OF THE BAIKAL EXPERIMENT ON OBSERVATIONS OF MACROSCOPIC NONLOCAL CORRELATIONS IN REVERSE TIME*

S.M. KOROTAEV, V.O. SERDYUK, E.O. KIKTENKO

*Geoelectromagnetic Research Centre of Schmidt Institute of Physics of the Earth, Russian Academy of Sciences, Troitsk, Moscow, 142190 Russia*

N.M. BUDNEV

*Institute of Applied Physics of Irkutsk State University, Irkutsk 664003 Russia*

J.V. GOROHOV

*Pushkov Institute of Terrestrial Magnetism, Ionosphere and Radio Wave Propagation, Troitsk, Moscow, 142190 Russia*

Although the general theory macroscopic quantum entanglement of is still in its infancy, consideration of the matter in the framework of action-at-a distance electrodynamics predicts for the random dissipative processes observability of the advanced nonlocal correlations (time reversal causality). These correlations were really revealed in our previous experiments with some large-scale heliogeophysical processes as the source ones and the lab detectors as the probe ones. Recently a new experiment has been performing on the base of Baikal Deep Water Neutrino Observatory. The thick water layer is an excellent shield against any local impacts on the detectors. The first annual series 2012/2013 has demonstrated that detector signals respond to the heliogeophysical (external) processes and causal connection of the signals directed downwards: from the Earth surface to the Baikal floor. But this nonlocal connection proved to be in reverse time. In addition advanced nonlocal correlation of the detector signal with the regional source-process: the random component of hydrological activity in the upper layer was revealed and the possibility of its forecast on nonlocal correlations was demonstrated. But the strongest macroscopic nonlocal correlations are observed at extremely low frequencies, that is at periods of several months. Therefore the above results should be verified in a longer experiment. We verify them by data of the second annual series 2013/2014 of the Baikal experiment. All the results have been confirmed, although some quantitative parameters of correlations and time reversal causal links turned out different due to nonstationarity of the source-processes. A new result is displaying of the advanced response of nonlocal correlation detector to the earthquake. This opens up the prospect of the earthquake forecast on the new physical principle, although further confirmation in the next events is certainly needed. The continuation of the Baikal experiment with expanded program is burning.

## 1. Introduction

According to the principle of weak causality [1], for the unknown quantum states (or, in other terms, for the random processes) advanced nonlocal correlations through a timelike interval and hence time reversal causality are possible. Such a time reversal approach provides uniform interpretation of the experiments on observation of the advanced correlation in teleportation [2], in entanglement swapping (which is teleportation of entanglement) [3, 4], and in some simple (basically) experiments with quantum interference [5]. On the other hand, more and more attention is drawn to the problem of macroscopic nonlocal correlations (e.g., [4]). Although strict theory of macroscopic nonlocality intriguing phenomenon is still in its infancy, it is known that rather universal means to create and maintain natural entanglement, including the macroscopic scale at the finite temperature, is dissipation [7-11]. It bridges the recent research with the early works of Kozyrev, who likely was the first to observe macroscopic entanglement of the dissipative processes [12]. Further such strange time reversal causality might be found in the nonlocally correlated random dissipative processes at macroscopic level too. Indeed in the large series of experiments advanced correlations between large-scale natural random dissipative processes and probe-processes in the detectors were revealed, nonlocal nature of the correlation had been proven by violation of Bell-like inequality, and time reversal causality had been proven by formal causal analysis [13-21]. Moreover the method of forecasting of the large-scale heliogeophysical processes on macroscopic nonlocal correlations had suggested and successfully tested [22-26].

* This work is supported by RFBR (grants 12-05-00001, 14-02-10003, 13-02-10002 and 14-45-04043) and CGPRF (grant SP-961.2013.5).

However these experiments are rather difficult because all known local impacts influencing the probe process in the detector, namely, temperature, pressure, electromagnetic field field, etc. must be excluded technically or/and mathematically. Therewith the strongest macroscopic nonlocal correlations are observed at extremely low frequencies, (at periods of several months), that is the long-terms experiments are necessary. Therefore we have to maintain very stable conditions in the nonlocal correlation detectors during very long time. It is extremely difficult in a usual laboratory. For example a passive thermostatting works well only at relatively short periods (while an active one is unfit because it brings the interference).

To overcome these difficulties a new experiment has begun on the base of Baikal Deep Water Neutrino Observatory. Baikal is the deepest lake in the World and its thick water layer is an excellent shield against the classical local impacts. In particular, the temperature near the floor is constant up to 0.01 *K*. The first annual series 2012/2013 has demonstrated that detector signals respond to the heliogeophysical (external) processes and causal connection of the signals directed downwards: from the Earth surface to Baikal floor. But this nonlocal connection proved to be in reverse time. In addition advanced nonlocal correlation of the detector signal with the regional source-process: the random component of hydrological activity in the upper layer was revealed and the possibility of its forecast on nonlocal correlations was demonstrated [27]. But since the typical periods are large, the experiment needs to be continued. In 2013/2014 the second annual measurement series with the modified setup has been obtained. In the current article we present the results of this second stage of the experiments and compare them with the first one. In addition a new result concerning the advanced response of nonlocal correlation detector to the earthquake is presented.

In Sec. 2 we consider shortly statement of the problem. The experiment is described in Sec. 3. The results are presented and discussed in Se. 4. We conclude in Sec. 4.

## 2. Statement of the problem

Theoretically quanta advanced correlations appear from different reasoning (e.g. [2, 5, 28]). We lean upon approach based on action-at-a-distance electrodynamics [1, 29, 30]. This theory considers the direct particle field as superposition of the retarded and advanced ones. The advanced field is unobservable and manifests itself only via radiation damping, which can be related with the entropy production [16, 26]. Any dissipative process is ultimately related with the radiation damping and therefore the advanced field (together with the retarded one) connects the dissipative processes. On the basis of all experimental data and theoretical considerations the following heuristic equation of macroscopic entanglement was suggested [14, 16, 19, 26]:

$$\dot{S}_d(t) = \sigma \int \frac{\dot{s}(\mathbf{r},t')}{|\mathbf{r}|} \delta\left(v^2 (t-t')^2 - \mathbf{r}^2\right) dV. \qquad (1)$$

It relates $\dot{S}_d(t)$ – the entropy production per particle in a probe process located in coordinate origin (that is a detector) with $\dot{s}(\mathbf{r},t')$ – the density of total entropy production in sources. Here $dV = v dt' d^3\mathbf{r}$ is 4-dimensional volume element of sources (analogous to that in special theory of relativity with $c \equiv v$), $\sigma$ is a cross section of transaction (it is of an atom order and goes to zero in the classical limit): $\sigma \approx \hbar^4 / m_e^2 e^4$, $m_e$ is the electron mass, $e$ is the elementary charge. The $\delta$-function shows that the processes are correlated with symmetric retardation and advancement. The propagation velocity $v$ for diffusion entanglement swapping can be very small. Accordingly, the retardation and advancement can be very large.

In general, Eq. (1) presents compact version of more intuitive empirical relation, which is directly suggested by all experiments carried out so far:

$$\dot{S}_d(t) = \frac{\sigma}{2} \int \frac{\dot{s}\left(\mathbf{r}, t+|\mathbf{r}|/v\right)}{\mathbf{r}^2} d^3\mathbf{r} + \frac{\sigma}{2} \int \frac{\dot{s}\left(\mathbf{r}, t-|\mathbf{r}|/v\right)}{\mathbf{r}^2} d^3\mathbf{r}\,. \qquad (2)$$

Mathematically eq. (1) and (2) are equivalent. Eq. (2) however expresses the proposed meaning more explicitly: a detector responds to an entropy production in source with symmetric retardation and advancement in time equal to $|\mathbf{r}|/v$. We expect both advanced and retarded responses for highly random source processes, for "manually" controlled sources one naturally expects only retarded response (see more details in [26]). Historically we got used to present the discussed equation in its former version due to compactness. In order to avoid possible confusions we would like to notice here also that in our previous publications (e.g. [13-27]) under $dV$ by default we were assuming 4-dimensional volume element similar to the cited above.

Also one implicit assumption in (1)-(2) is that the rate of entropy change is the same for source and detector. From empirical data this is not necessarily the case (see e.g. fig. 9 below - detector entropy growth caused presumably by the earthquake was lasting over many days). One may easily account such a rate difference simply by rescaling the time argument (1) by some constant, which may in turn depend on the setup. However, we do not introduce such a correction so far due to ambiguity of reasons of such a rate difference: it could be either a true inherent primary feature of elementary entropy exchange or just a secondary effect due to entropy "re-radiation" by ambient and intra-detector medium. And another important property of (1)-(2) is their ability to satisfy total entropy conservation during its transfer from a source to detector. One can easily see this by noticing that an absorbing particle is "viewed" by point entropy source under the solid angle $\Delta\Omega = \sigma / \mathbf{r}^2$. For the precise conservation an additional normalization factor $4\pi$ is actually needed. However, due to very high experimental uncertainties and setup dependence in the overall normalization of the r.h.s., we do not specify this normalization constant at such a detailed level yet.

Eq. (1) was tested theoretically (for the spin gas model [31]) [25, 26] and experimentally with electrode, photocathode and ion mobility detectors (under condition of small absorption by the intermediate medium) [16, 19, 26].

It should be noted that Eq. (1) does not take into account the absorption by the intermediate medium. Its influence, however, is very peculiar. Although the equations of action-at-a-distance electrodynamics are time symmetric, the fundamental time asymmetry is represented by the absorption efficiency asymmetry: the absorption of retarded field is perfect, while the absorption of advanced one must be imperfect [15, 20, 25, 30]. It leads to the fact, that level of advanced correlation through a screening medium may exceed the retarded one. In addition, if this correlation asymmetry is not too large, the third, apparent instantaneous (quasi-synchronous) correlation may exist as a result of advanced/retarded signal interference [1, 29].

So the experimental problem is to establish correlation between the entropy variations in the probe- and source-processes under condition of suppression of all classical local impacts. The detector based on spontaneous variations of self-potentials of weakly polarized electrodes in an electrolyte proved to be the most reliable one. The theory of the electrode detector starts from self-consistent solution of the entropy production in the liquid phase. The entropy of distribution can be expressed in terms of full contact potential. From here one can get the expression of the entropy variation in terms of potential difference between a couple of electrodes, which is the detector signal [13, 14, 17, 20, 26].

The most prominent achievement of the previous experiments [13-27] was reliable detection of advanced macroscopic nonlocal correlations and experimental proof of time reversal causality for the random processes. The mathematical tool for this proof is causal analysis [26, 32], which recently plays also important role in theoretical studies of quantum information problems. Although the considered phenomenon is quantum, but in the experiment we deal with the classical output of measuring device and we can use simpler classical causal analysis. Generally, classical causal analysis can be used instead of more complicated quantum one if the conditional entropies are non-negative. The kernel of the method is as follows. For any variables $X$ and $Y$ several parameters of their interrelation can be defined in terms of Shannon marginal $S(X)$, $S(Y)$ and conditional $S(X|Y)$, $S(Y|X)$ entropies. The most important are the independence functions:

$$i_{Y|X} = \frac{S(Y\,|\,X)}{S(Y)},\; i_{X|Y} = \frac{S(X\,|\,Y)}{S(X)},\; 0 \le i \le 1. \qquad (2)$$

Roughly saying, the independence functions behave inversely to module of correlation one. But they characterize one-way correlations, which are asymmetric for causally related variables. In addition they work equally well for linear or any nonlinear relations. Next the causality function $\gamma$ is considered:

$$\gamma = \frac{i_{Y|X}}{i_{X|Y}},\; 0 \le \gamma < \infty. \qquad (3)$$

By definition $X$ is the cause and $Y$ is the effect if $\gamma<1$. And inversely, $Y$ is the cause and $X$ is the effect of $\gamma>1$. On theoretical and plenty of experimental examples it had been shown that such a formal approach to causality did not contradict its intuitive understanding in the simple situations and could be fruitfully used in complicated ones. In terms of $\gamma$ the principle of classical causality is formulated as follows:

$$\gamma < 1 \Rightarrow \tau > 0,\ \gamma > 1 \Rightarrow \tau < 0,\ \gamma \to 1 \Rightarrow \tau \to 0, \quad (4)$$

where $\tau$ is time shift of *Y* relative to *X*. Only in case of nonlocal correlation one can observe violation of this principle. It is just the case of weak causality [1], which does not obey the combination of inequalities of axiom (4).

## 3. Experimental setup

The Baikal experiment has been carried out since 2012. The experiment aims, first, study of nonlocal correlation between the detectors at different horizons in the lake and spaced at 4200 *km* lab detector in Troitsk, and second, study of correlations of detector signals with the natural dissipative source-processes with big random component.

The experimental setup is deployed at site φ = 51°.721 N, λ = 104°.416 E. The site depth is 1367 *m*. The first of the annual series 2012/2013 was, essentially, a test experiment, so for the second annual series 2013/2014 the setup configuration was slightly modified to improve setup's reliability and to increase detector spacing. For this reason we can not obtain two-year series yet and in Sec. 3 we consider results of the second one-year series in a brief comparison with the corresponding results of the first one, presented in detail in Ref. [27].

In Figure1 the scheme of Baikal Deep Water Setup is shown. The bottom detector is set at the depth 1337 *m*, the top one is set at the depth 47 *m*. Both the detectors represent a couple of high quality weakly polarized *AgClAg* electrodes HD-5.519.00 with practically zero separation. These electrodes were originally designed for high precision measurements of the weak electric fields in the ocean, and they are best in the World by their self-potential insensitivity to the environmental conditions. The signal are measured and stored in the electronics unit set at the depth 20 *m*. The sampling rate is 10 *s*. The calibration and zero control are done automatically daily. The relative error of measurements is less than 0.01%. In addition, the electronics unit contains the temperature and acceleration sensors. The setup is fixed by the heavy anchor on the floor and by the drowned buoy at the depth 15 *m*.

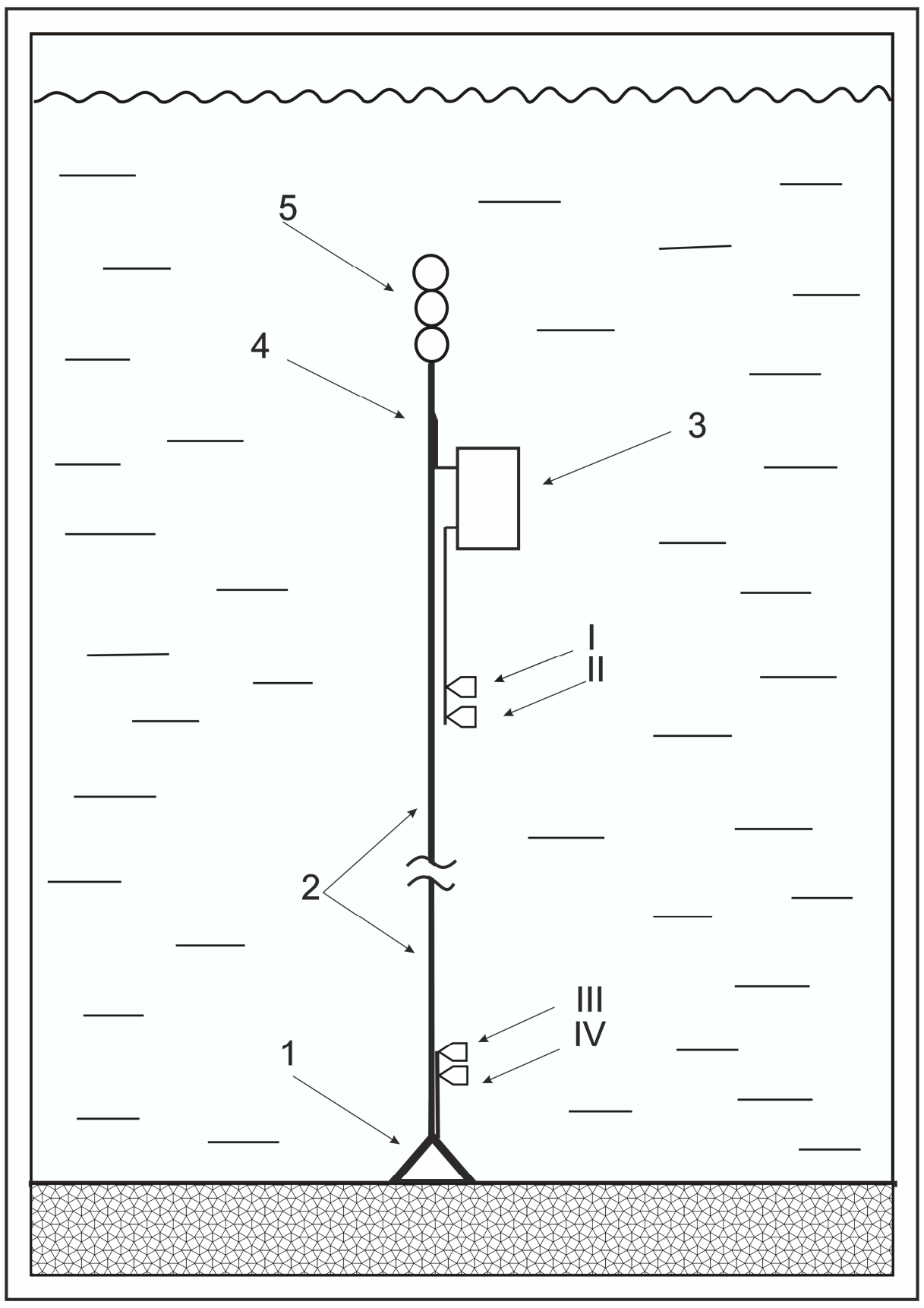

Figure 1. Baikal Deep Water Setup (1 – anchor; 2 – cable; 3 – electronics unit, acceleration and temperature sensors; 4 – buoy rope; 5 – buoy; I, II – top electrode detector; III, IV – bottom electrode detector).

The setup is designed to be operated autonomically for a year. For the second annual series it was installed from the ice in March, 2013. In March, 2014 the setup was lifted on the ice for data reading and battery changing and then it was installed again for the next year. It is known that the strongest macroscopic nonlocal correlations are observed at the long periods. Therefore our experiment is planed for several years with the possible expansion of the program.

Data were processed by the methods of spectral, causal and also usual correlation analysis.

## 4. Results

From the classical standpoint the detector signals must be uncorrelated noises. But it is not the case. In Figure 2 the normalized amplitude spectra of the bottom detector $U_b$, top one $U_t$ and far distant Troitsk lab one $U_l$ are presented. The period range is from 10 to 220 days (*d*). It is seen that at the longest periods the spectra are similar. We observe in all the detectors the solar intermittent variation at about 140 *d* (this is strong solar X-ray variation with big random component [33], which is known by satellite data only, because X-ray radiation is fully absorbed in the upper atmosphere). But another solar variation with small random component (the split maxima around period of solar rotation 27 *d*) is considerable only in the spectrum of the detector on the Earth surface $U_l$. It is also seen that spectrum of $U_b$ more exactly corresponds to close $U_t$ one than to distant *Ul* one. The best all spectra similarity is observed at the period range 145 > *T* > 46 *d*. In 2012/2013 series [27] the spectra were akin (of course, not in detail), the range of similarity was shifted to the longer periods (T > 77 *d*).

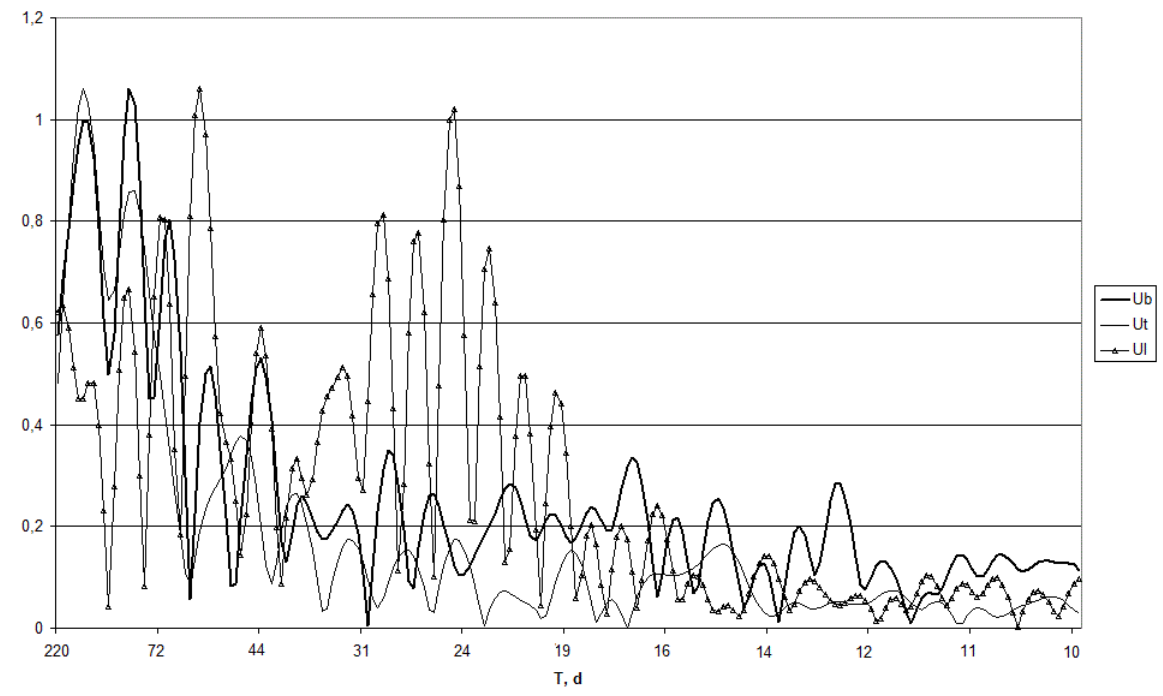

Figure 2. Normalized amplitude spectra of the signals of bottom detector $U_b$, top one $U_t$ and lab one $U_l$.

For causal and correlation analysis we used broadband filtered data (145 > *T* > 46 *d*). Hereafter the relative errors of $\gamma$ and $i_{X|Y}$ estimations are less than 10%. In Figure 3 the results for the bottom $U_b$ and top $U_t$ detectors are presented. $\gamma > 1$ that is $U_t$ is the cause and $U_b$ is the effect. At $\tau > 0$ we observe classically forbidden time reversal causality. It is just weak causality allowed only for the entangled states. The highest maximum of $\gamma = 5.2$ is at advancement 21 *d* . Each maximum of causality $\gamma$ corresponds to minimum of independence $i_{X|Y}$. The deepest minimum of $i_{X|Y}$. = 0.075 is at advancement 11 *d*, and it corresponds to maximal negative correlation function $r = -0.98$ (but the latter is equal by module to the retarded one). There is approximate symmetry between the retardation (– 24 *d*) and advancement (21 *d*) in the causal link $U_t \rightarrow U_b$, however the shapes of $\gamma$-maxima differ considerably.

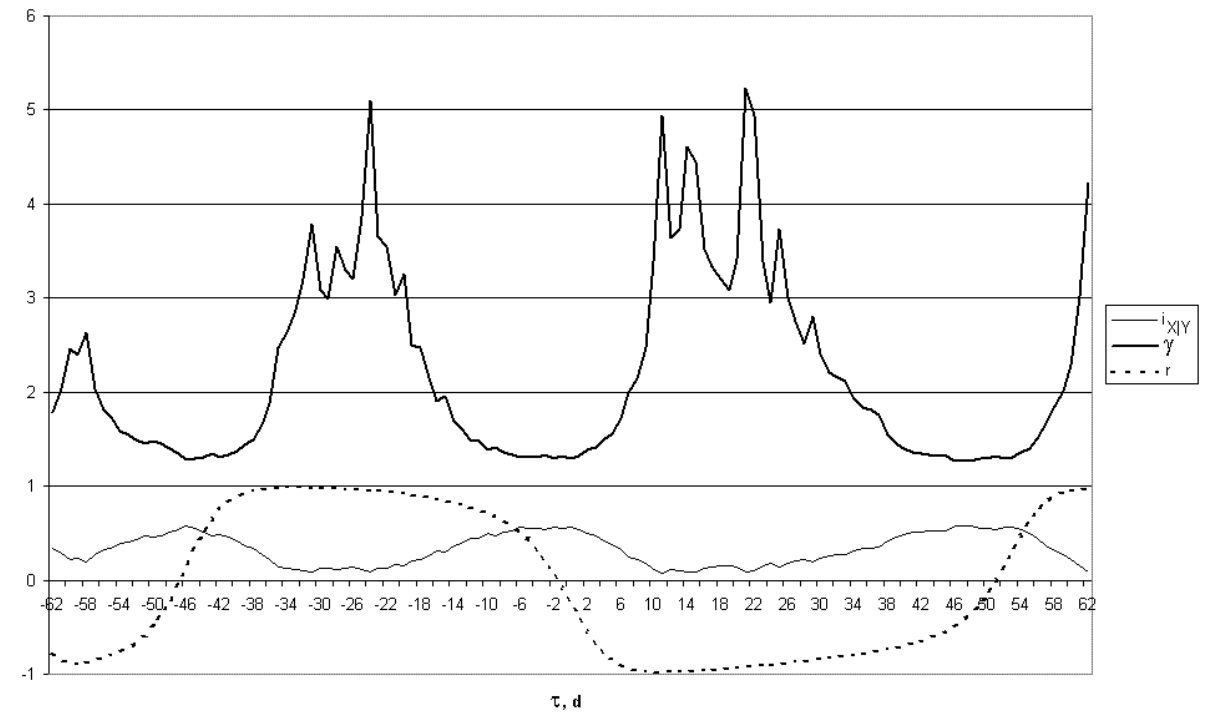

Figure 3. Causal and correlation analysis of $U_b$ (*X*) and $U_t$ (*Y*). $\tau < 0$ corresponds to retardation of $U_b$ relative $U_t$, $\tau > 0$ – to advancement.

In Figure 4 the result of the same analysis of the top detector $U_t$ and the distant one $U_l$ is presented. We observe $\gamma > 1$ that is $U_l$ is a cause with respect to $U_t$. The shapes of retarded and advanced $\gamma$-maxima are similar and as a result there is a quasi-synchronous $\gamma$-maximum ($\tau \approx 0$). But again the highest maximum of $\gamma = 1.4$ is advanced ($\tau = 29$ *d*). The strongest correlation $r = -0.87 \pm 0.01$ is at larger advancement $\tau = 56$ *d* (the correlation function reflects only linear relation; therefore positions of its extreme can considerably differ).

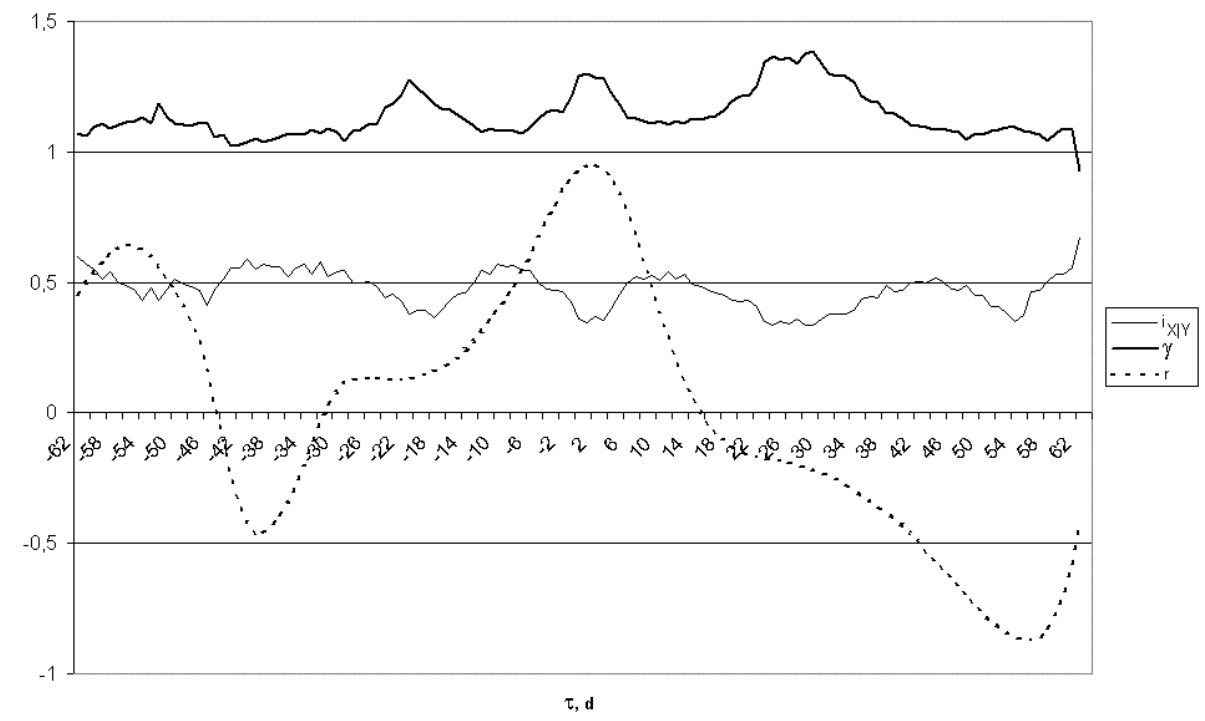

Figure 4. Causal and correlation analysis of $U_t$ (*X*) and $U_l$ (*Y*). $\tau < 0$ corresponds to retardation of $U_t$ relative $U_l$, $\tau > 0$ – to advancement.

And in Figure 5 causal analysis of the bottom detector $U_b$ and distant one $U_l$ is presented. Again $U_l$ is a cause with respect to $U_b$, but prevailing of time reversal causality is expressed much stronger: the advanced $\gamma$-maximum is 2.1 times as higher as the retarded one. The highest $\max\gamma = 6.2$ is at advancement 24 *d*. It exactly corresponds to the deepest $\min i_{X|Y} = 0.078$. There is also an obvious considerable time asymmetry in advanced/retarded time shifts of the all extreme. Therefore there is no a quasi-synchronous $\gamma$-maximum (although the modules of advanced and retarded correlation functions are practically equal (0.95)).

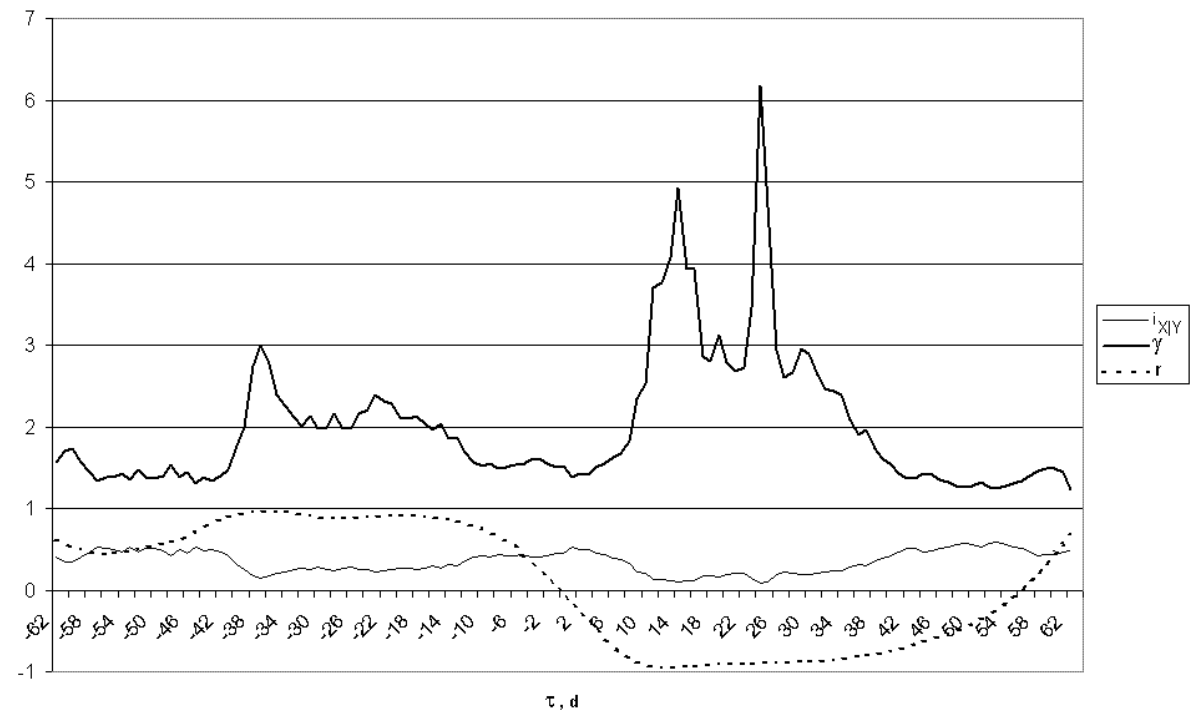

Figure 5. Causal and correlation analysis of $U_b$ ($X$) and $U_l$ ($Y$). $\tau < 0$ corresponds to retardation of $U_b$ relative $U_l$,, $\tau > 0$ – to advancement.

Thus we may conclude that by data of three detectors the causal connection is directed downwards, from the Earth surface to the lake floor: $U_l \to U_t$, $U_l \to U_b$, $U_t \to U_b$ It is quite natural for the external heliogeophysical source-processes which by data of all previous studies [13-26] are the main processes entangled with spontaneous probe-processes in the detectors on the Earth surface, but this causality is time reversal: the effects appear before the causes! The same conclusion was made by 2012/2013 [27], there is only some quantitative difference owing to nonstationarity. The qualitative peculiarities are the same too, e.g. the link not separated by the water layer ($U_l \to U_t$) demonstrates the relatively lowest $\gamma$.

Note, that that the links $U_l \to U_t$, $U_l \to U_b$, $U_t \to U_b$ do not constitute a causal chain $U_l \to U_t \to U_b$: the values of $\tau$ are subadditive, because the connection is nonlocal.

Among the closer, regional dissipative source-processes with big random component the most obvious are water temperature variations. These variations occur in the rather thin subsurface layer (active layer in hydrological terms). So the temperature measured by the setup at the depth 20 *m* is just representative for them. Of course, we must keep in mind a possible classical local influence of the temperature variations on the electrode self-potentials according to the temperature coefficient of given electrode pairs (for our detectors it equals 0.04 *mV/K*). As the temperature amplitude strongly decays with the depth (the $U_t$ detector is set 27 *m* deeper than the *t* sensor) the spectral amplitude ratio $U_t$ / *t* must be less than this value. In regard to *Ub* placed near Baikal Lake floor, it is absolutely reliable protected against classical temperature influence, while nonlocal influence may be suppressed too due to the screening by very thick water layer (greater than 1300 *m*).

In Figure 6 the amplitude spectra (in absolute units) of both the detectors and temperature *t* in the active layer are presented. It is seen that at the long periods there is a certain similarity of the spectra $U_t$. and *t*. But their amplitude ratio proves to be much greater (of order 1 *mV/K*) than upper bound of a local influence (0.04 *mV/K*). It may be explained only by some nonlocal correlation.

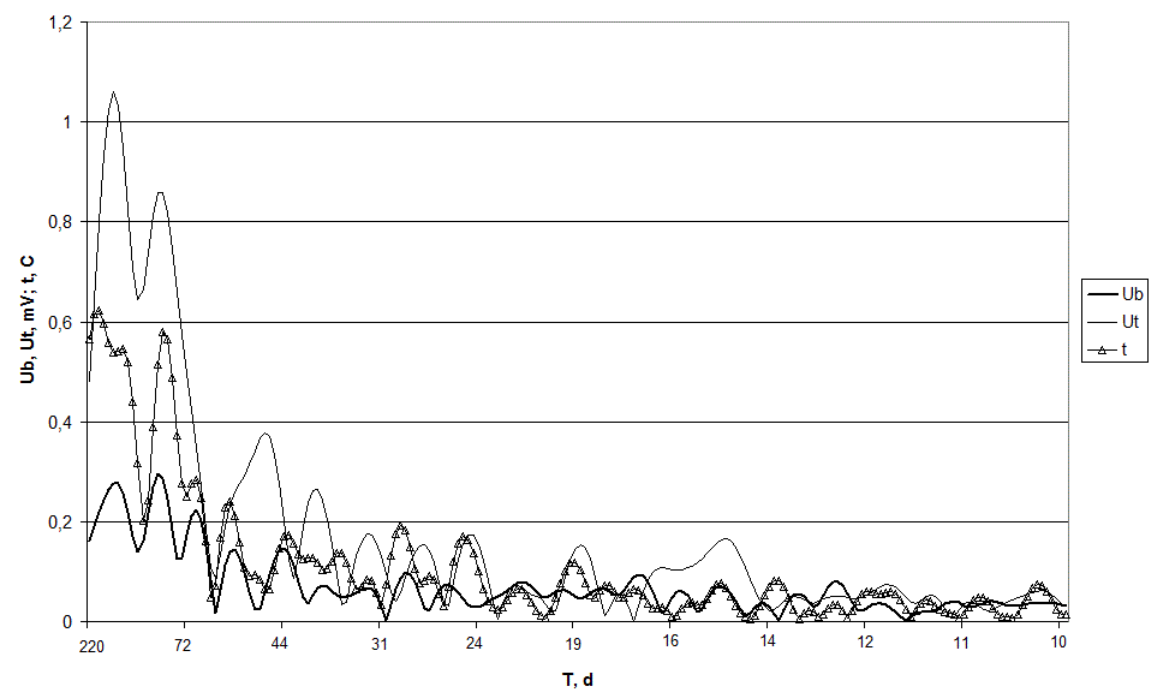

Figure 6. Amplitude spectra of the bottom detector $U_b$, top one $U_t$ and temperature *t* at the depth 20 *m*.

Consider the results of causal and correlation analysis of $U_t$ and *t* (Figure 7) with the same broadband filtration. We observe that *t* is a cause with respect to $U_t$ with three (retarded, quasi-synchronous, and advanced) $\gamma$-maxima with exactly corresponding $i_{X|Y}$-minima. The retarded and advanced $\gamma$-maxima are similar that implies appearance of the intermediate quasi-synchronous one. But again the highest $\max\gamma = 1.5$ and the deepest $\min i_{X|Y} = 0.24$ are advanced ($\tau$ = 20 *d*). The corresponding advanced maximum of negative correlation function $r = -0.97$ is exactly at the same $\tau$ (although *r* equals by module the retarded one). It is just manifestation of advanced nonlocal connection $t \to U_t$.

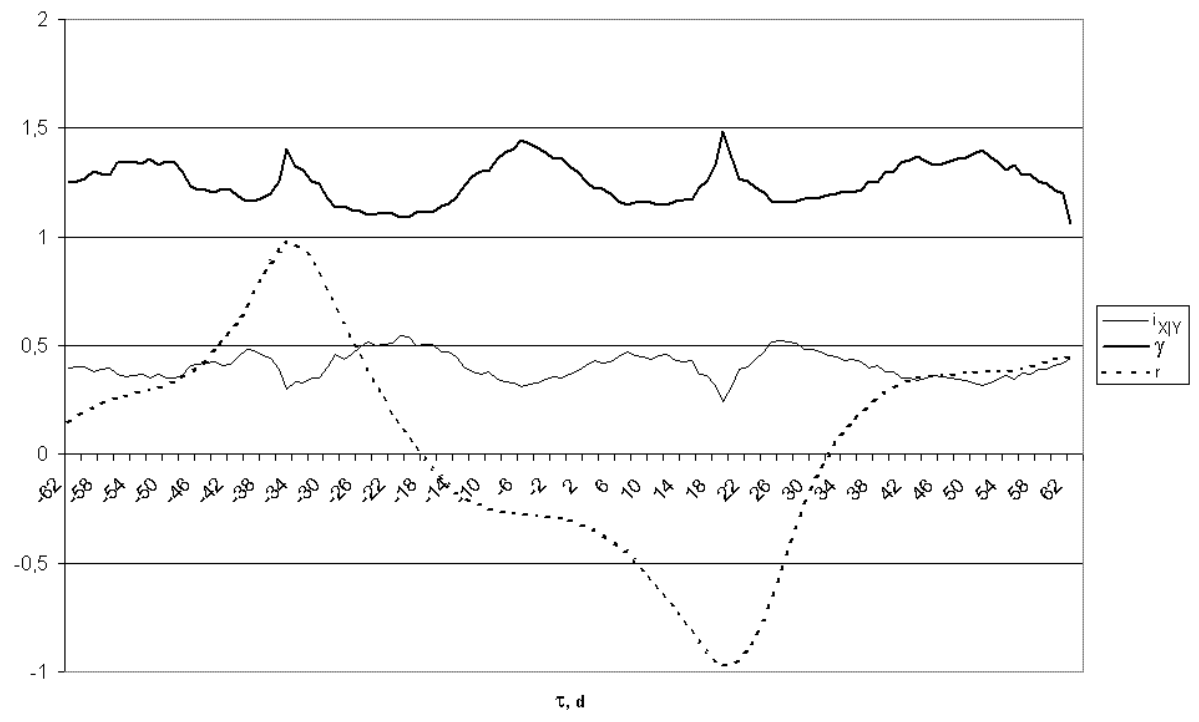

Figure 7. Causal and correlation analysis of $U_t$ ($X$) and $t$ ($Y$). $\tau < 0$ corresponds to retardation of $U_t$ relative $t$,, $\tau > 0$ – to advancement.

We have applied to these data the forecasting algorithm based on computation of current (sliding) regression. This algorithm needs rather long training interval; hence we could test the forecast only by relatively short segment of the time series. The result is presented in Figure 8. The forecast curve showed in this figure is obtained by means of day by day forecast with fixed advancement $\tau$ = 20 $d$. The accuracy of the forecast is acceptable for all practical purposes.

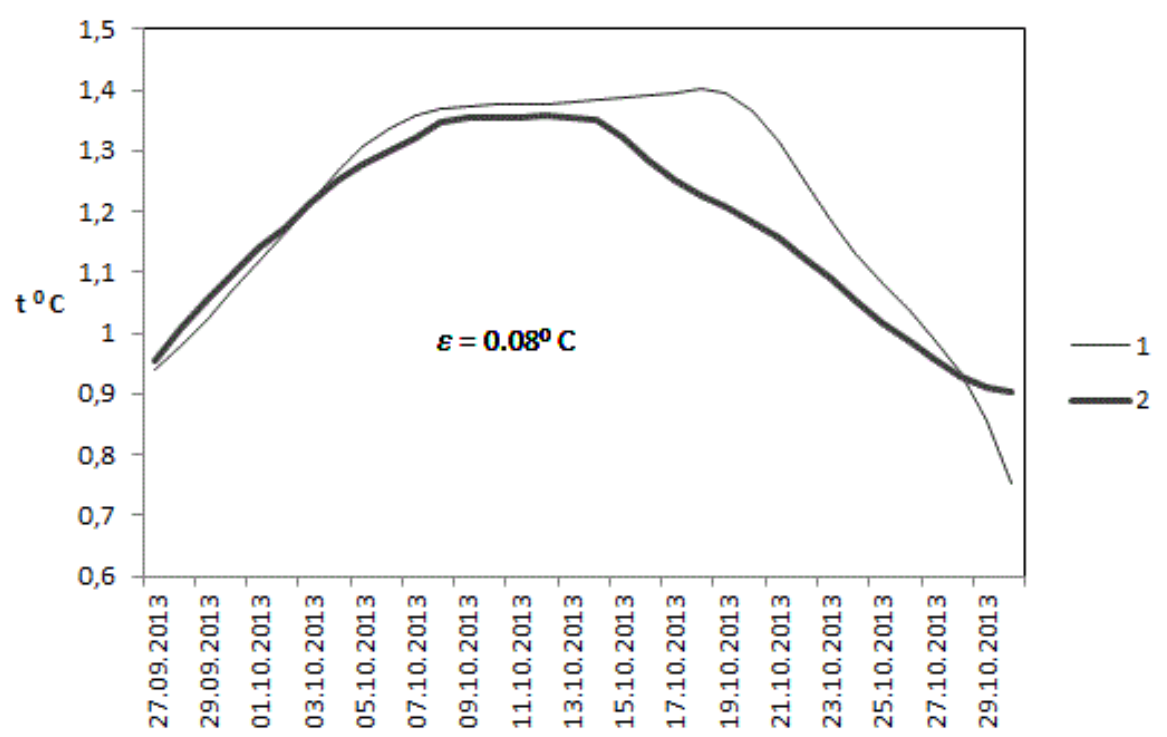

Figure 8. The forecast test segment of the active layer temperature with advancement 20 days (1) as compared to the factual one (2). The ε is the standard deviation of forecast and actual curves.

The results concerning nonlocal connection $t \rightarrow U_t$ described above is qualitatively the same as in the previous experiment, the quantitative difference is in the greater advancement $\tau$ = 45 $d$ [27]. It is unclear is it effect of nonstationarity only, or deeper position of $U_l$ detector (52 m) in the previous setup configuration.

Other regional very powerful random dissipative source-processes are the earthquakes. Although the Baikal Rift is tectonically very active, the earthquakes are rare events in the achieved scale of duration of the experiment. At last, quite a powerful earthquake (with magnitude $M$ = 5.6) occurred on December 22, 2013 with epicenter at φ = 53°.7 N, λ = 91°.4 (890 $km$ from our observation site) and with depth of hypocenter 15 $km$.

In the bottom detector signal $U_b$ this event can be seen without any processing (Figure 9) as a single for the whole year short-term disturbance in the form of the characteristic triple burst (advanced – quasi-synchronous – retarded). The disturbance begins 12 days before the event, reaches the advanced (left one in Figure 9) maximum about 6 days before the event and, then, having the quasi-synchronous one (central) reaches the retarded one (right) about 14 days after the event, and at last completely disappeared 25 days after the event. The voltage excursion of with regard to trend equals 2.7 $mV$. For comparison the temperature $t$ as the main potential interference is also shown in Figure 9, and it is seen that $U_b$ and $t$ are completely unrelated.

Contrary to $U_b$, the annual record of $U_t$ is rather indented due to the processes in the close active layer and above. Nevertheless, near the moment of $U_b$ advanced maximum we observe the unprecedented sharp irreversible change of $U_t$. The voltage excursion equals 2.7 $mV$ too. Most probably it is some another sort of an advanced response, which we can not realize yet.

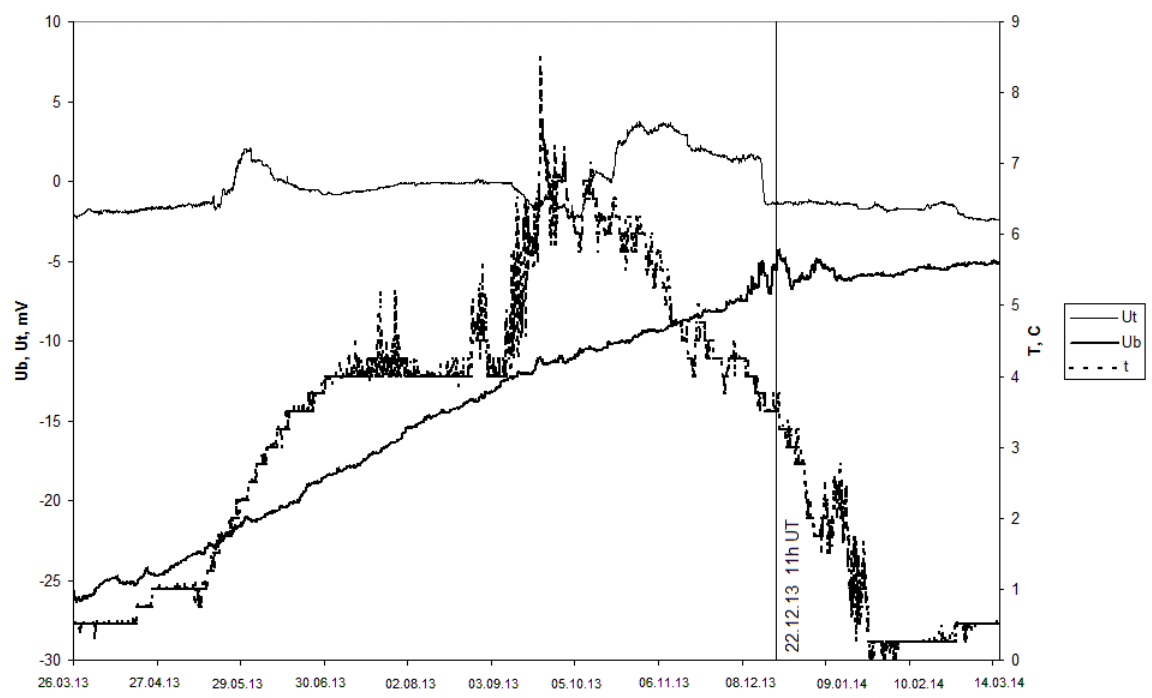

Figure 9. Hourly average values of signals of the top $U_t$ and bottom $U_b$ detectors ($t$ is active layer temperature). The vertical line is the earthquake moment.

In any event, the signal disturbance in the seabed bottom detector $U_b$ is more understandable, practically noise-free and the phenomenon picture is in line with previous macroscopic nonlocality research. This opens up the prospect of the earthquake forecast on the new physical principle, although further confirmation in the next events is certainly needed.

## 5. Conclusion

The long-term Baikal Deep Water Experiment on study of macroscopic entanglement and related phenomena of advanced nonlocal correlations in reverse time is under way; the second annual data series has been obtained. The experiment includes measurements with three nonlocal correlation detectors at the two horizons in the Baikal Lake and at the distant land. Nonlocal correlations between detector signals and between them and large-scale random geophysical source-processes are studied.

The detector signals proved to be nonlocally causally connected with approximately symmetric retardation and advancement. Both advanced and retarded nonlocal correlations correspond to the same direction of causality. Therewith time reversal causality prevails over usual time respecting one. It is the most prominent property of macroscopic entanglement and manifestation of quantum principle of weak causality.

Study of two regional random source-processes: hydrological activity in the Baikal Lake upper layer, and the earthquake, has revealed advanced detector signal responses, which can be used for the practical forecasts.

## References

1. J. G. Cramer, Phys. *Rev.* **D22**, 362 (1980).
2. M. Laforest, J. Baugh and R. Laflamme, *Phys. Rev*. **A73**, 032323 (2006).
3. X.-S. Ma, S. Zotter, J. Kofler, R. Ursin, T. Jennewien, Č. Brukner and A. Zeilinger, *Nature Physics* **8,** 479 (2012).
4. E. Megidish, A. Halevy, T. Shacham, T. Dvir, L. Dovrat and H.S. Eisenberg, *Phys. Rev. Lett.* **110**, 210403 (2013).
5. A. Danan, D. Farfurnik, S. Bar-Ad and L. Vaidman, *Phys. Rev. Lett.* **111** 240402 (2013).
6. M. Yu. Terri, K. R. Brown and I. I. Chuang, *Phys. Rev.* **A71**, 032341 (2005).
7. M. B. Plenio and S.F Huelga, *Phys. Rev. Lett.* **88**, 197901 (2002).
8. M. S. Kim, J. Lee, D. Ahn and P.L. Knight, *Phys. Rev.* **A65**, 040101 (2002).
9. D. Braun, *Phys. Rev. Lett.* **89**, 277901 (2002).
10. F. Benatti, R. Floreanini and M. Piani, *Phys. Rev. Lett.* **91**, 070402 (2003).
11. T. Choi and H. J. Lee, *Phys. Rev.* **A76**, 012308 (2007).
12. N. A. Kozyrev, “On the possibility of experimental investigation of the properties of time”, in *Time in Science and Philosophy*, edited by J. Zeman, Prague: Academia, 1971, pp. 111-132.
13. S. M. Korotaev, A. N. Morozov, V. O. Serdyuk and M. O. Sorokin, *Russian Phys. J.*, **45**, (5), 3 (2002).
14. S. M. Korotaev, A. N. Morozov, V. O. Serdyuk and J.V. Gorohov, “Experimental evidence of nonlocal transaction in reverse time”, in: *Physical Interpretation of Relativity Theory*, edited by M. C. Duffy, V. O. Gladyshev and A.N. Morozov, Moscow, Liverpool, Sunderland: BMSTU PH, 2003, pp. 200-212.
15. S. M. Korotaev, V. O. Serdyuk, V. I. Nalivaiko, A.V. Novysh, S. P. Gaidash, Yu. V. Gorokhov, S. A. Pulinets. and Kh. D. Kanonidi., *Phys. of Wave Phenomena* **11**, 46 (2003).
16. S. M. Korotaev, V. O. Serdyuk, J.V. Gorohov, S. A. Pulinets and V. A. Machinin, *Frontier Perspectives* **13** (1), 41 (2004)
17. S. M. Korotaev, A. N. Morozov, V. O. Serdyuk, J.V. Gorohov and V. A. Machinin, *NeuroQuantology* **3**, 275 (2005).
18. S. M. Korotaev, A. N. Morozov, V. O. Serdyuk V. I. Nalivaiko, A.V. Novysh, S. P. Gaidash, Yu. V. Gorokhov, S. A. Pulinets. and Kh. D. Kanonidi, *Vestnik J. BMSTU* **SI**, 173-(2005).
19. S. M. Korotaev, A. N. Morozov, V. O. Serdyuk, J.V. Gorohov and V. A. Machinin, “Experimental study of advanced nonlocal correlation of large-scale dissipative processes”, In: *Physical Interpretation of Relativity Theory*, edited by M. C. Duffy, V. O. Gladyshev A.N. Morozov and P. Rowlands Moscow, Liverpool, Sunderland: BMSTU PH, 2005, pp. 209-231.
20. S. M. Korotaev, *Int. J. of Computing Anticipatory Systems* **17**, 61 (2006).
21. S. M. Korotaev, A. N. Morozov, V. O. Serdyuk, J. V. Gorohov, V. A. Machinin and B. P. Filippov, *Russian Physics Journal* **50**, 333 (2007).
22. S. M. Korotaev, V. O. Serdyuk and J. V. Gorohov, “Signals in reverse time from heliogeophysical random processes and their employment for the long-term forecast”, in: *Physical Interpretation of Relativity Theory*, edited by M. C. Duffy, V. O. Gladyshev A.N. Morozov and P. Rowlands Moscow, Liverpool, Sunderland: BMSTU PH, 2007, pp. 222-230.
23. S. M. Korotaev, V. O. Serdyuk and J.V. Gorohov, *Hadronic Journal* **30**, 39 (2007).

24. S. M. Korotaev, V. O. Serdyuk and J. V. Gorohov, *Doklady Earth Sciences* **415A**, 975-978 (2007).
25. S. M. Korotaev and V. O. Serdyuk, *Int. J. of Computing Anticipatory Systems* **20**, 31 (2008,).
26. S. M. Korotaev, *Causality and Reversibility in Irreversible Time*, Scientific Research Publishing, 2011.
27. S. M Korotaev, N. M. Budnev, V. O. Serdyuk, J. V.Gorohov, E. O Kiktenko., V. L. Zurbanov, R.R. Mirgazov, V. B Buzin. and A. V. Novysh, "Preliminary results of the Baikal experiment on observations of macroscopic nonlocal correlations in reverse time", in: *Physical Interpretations of Relativity Theory*, edited by V. O. Gladyshev A.N. Morozov and P. Rowlands Moscow, Liverpool, Sunderland: BMSTU PH, 2013, pp. 141-151.
28. E. A. Rauscher and R. L. Amoroso, *Int. J. of Computing Anticipatory Systems* 22, 370 (2008).
29. J. G. Cramer , *Rev. Mod. Phys.* **58**, 647 (1986).
30. F. Hoyle and J. V. Narlikar, *Rev. Mod. Phys.* **67**, 113-156 (1995).
31. J. Calsamiglia, L. Hartmann, W. Dür and H.-J. Briegel, *Phys. Rev. Lett*. **95**, 180502 (2005).
32. S. M. Korotaev and E. O. Kiktenko, *AIP Conference Proceedings* **1316**, 295 (2010).
33. E. Reiger, G. H. Share and D. G. Forrest, *Nature*. **312**, 625 (1984).